# Equibit: A Peer-to-Peer Electronic Equity System

**Brent Kievit-Kylar**
brent@equibit.org
Software Engineer
Cambridge, MA
USA

**Chris Horlacher**
chris@equibit.org
Chartered Accountant
Toronto, ON
Canada

**Marc Godard**
marc@equibit.org
Full Stack Developer
Toronto, ON
Canada

**Christian Saucier**
christian@equibit.org
Software Engineer
Raleigh, NC
USA

equibit.org

**Abstract:** The registration, transfer, clearing and settlement of equities represents a significant part of economic activity currently underserved by modern technological innovation. In addition, recent events have revealed problems of transparency, inviting public criticism and scrutiny from regulatory authorities. A peer-to-peer platform facilitating the creation and exchange of directly registered shares represents a shift in equity markets towards more efficient, transparent operations as well as greater accessibility to the investing public and issuing companies. Blockchain technology solves the problem of transaction processing and clearing but the fungibility of their units pose a challenge in identifying the issuers and holders of specific equities. Furthermore, as the issuers are in a constant state of flux the benefits of a decentralized network are lost if a central manager is required to cancel equities from companies that no longer exist. We propose a solution to these problems using digital signatures, based on blockchain technology.

## Introduction

Equity represents one of the largest classes of investments, typically controlled by central depositories and stock transfer agents. Despite its size, the overall equity market is opaque and market actors must rely on these intermediaries to raise capital, conduct trades, and accurately manage their shareholder registers. Extensive documentation is required in order to ensure that purchased units are authentic. Furthermore, issuers or their agents have to maintain a register of investor's identities and addresses in order to distribute earnings and collect votes on investor resolutions. This requires issuers to collect extensive information on their investors and incur significant costs in order to issue and transfer shares, distribute earnings and poll their investors.

What is needed is an electronic equity system allowing issuers to create, disseminate and maintain equity across a broad base of investors without the need for onerous recordkeeping and intermediaries.

To move away from central authority based systems, we must embrace distributed trust. The movement toward distributed trust systems has been building for years, growing out of the social movement facilitated by a connected world.[1,2] Its advantages include increased flexibility, lack of single point failure, openness and cost.

These social trust systems have recently been making waves in the domain of economics with the rise to prominence of digital currencies. Systems like Bitcoin provide a distributed trust system for value exchange.[3] These cryptocurrencies are designed around a public ledger protected by cryptographic primitives that allow individuals to move value from one user to another securely, insofar as value cannot be double spent, nor given if it is not owned.

The same features that allow bitcoin to be an effective monetary representation, can also easily handle the equity dissemination process and maintain a register of investors. With additional modifications it could also be made to allow for the creation of uniquely identifiable private shares within the blockchain itself, as well as distribute earnings to and collect votes from the relevant investors.

The following paper describes a blockchain able to be used as a registration and transfer vessel for equity instead of a pure digital currency. We describe the cost-benefit, trust, and flexibility of this new system as well as the communications and trust systems needed in order to make it an adequate platform for conducting investor relations and to serve as a secondary market.

## Equity Creation, Issuance and Cancellation

When a network node discovers new equibits the units contain a null issuer information field signifying that the unit is presently in a null state. This null state is the equivalent of a blank stock certificate, which would then be signed and sealed by an offering corporation and delivered to the investor. The Equibit system digitally models these same operations. Often this blank stock in itself has very little value but, as we will discuss later, there is a limited amount of these null units in the system. These new equibits enter the market and may be traded from one user to another as any other equibit and will likely carry some nominal value due to their usefulness in becoming issuer authorized in the future.

At some point this equibit will be transferred to a user in the network that has the need to authorize and issue shares of their own. The new user creates a transaction that cryptographically signs the null equibits using the user's private key. By posting a signature in the issuer information field, the new user announces to the network that these units are now a representation of equity in their particular business. At that point they may then transfer the equibit to another user or hold on to it. The issuer field of an authorized equibit would, at a minimum, contain the following information:

**Company Name:** Legal name of the issuer.

**Company Domicile:** Country/Jurisdiction where the company is registered.

**Security Name:** Name of the issuance.

**Security Type:** Signifies the legal form of the equity units; Common Shares, Preferred Shares, Trust Units, Fund Units, Partnership Units, etc.

**Restriction Level:** Indicates the level of transfer restrictions there are on a particular security. See the section on Trading Passports for more information.

Unlike traditional cryptocurrency, the value of an individual equibit comes from the issuer field and the presumed value that issuer provides. Thus, not all equibits are equal. Also, the real life identity of the issuers becomes an important factor in the value of the equibit, as compared to the relative anonymity desired by most cryptocurrencies. Fortunately, the prominence of public key encryption algorithms already requires these entities to have the infrastructure to provably distribute public keys and identification. Thus, if a public key can be trusted to belong to a given entity, then that entity can prove to any user that a particular transaction was performed by them, because no other entity could have the relevant private key. While the identities of the issuers must be resolved, the purchasers can choose whether or not to maintain the same level of anonymity available to traditional bitcoin users.[4]

The status of all equibits in the system is a matter of public record due to the very nature of the blockchain. Therefore, it will be easy for market participants to take note of the sum total of equibits a particular issuer has authorized, how many have left the originating address, as well as how many are still retained by the issuer for future offerings. This is important information for investors as they desire knowledge of what potential dilution of their investment they could experience in the future. They would also immediately know when an existing issuer authorizes new units by adding to an existing issuance.

Verifying the authenticity of units received is also an important part of the system. This can be accomplished by tracing the chain of transactions of the equibit units held at a particular address back to the address the authorization originated at. If the addresses and signatures match then the unit is authentic and should be accepted at its market value. If the signatures do not match then the unit is a forgery and should be immediately returned to its null state to avoid contaminating the Equibit network. Notice that equibits do not fall out of the system, they are always recycled.

Cancelling equibits is a simple matter of the holder overwriting the issuer field with the system default, null value. This restores that unit back to its blank state and it can then be authorized by a new issuer. We must allow anyone to be able to strip issuer's authorization information and recycle the units they possess, otherwise all of the equibits will eventually be used up. We cannot count on individuals or companies to continually scrub the system of worthless units without the proper incentive.

By building in a system of natural attrition we create a supply/demand dynamic between the authorized and blank units. This provides the mechanism through which the system will continually evolve with new issuers coming to market and terminated or declining business ventures being scrubbed from the system. As the maximum supply of blank units is fixed and determinable they will carry a certain value ascribed to them by users requiring them to pay transaction fees, as well as incoming businesses wanting to issue equity. They will be in competition with the existing issuers who want to maintain their own equity base by providing value to their investors. Thus if an existing issuer's value drops below the value of an equivalent number of blank equibits they run the risk of their investors recycling the equibits back to the more valuable blank units, which can be sold to new issuers. This also provides investors with a level of assurance that their investments will never go to zero, as they will always be able to recover the equivalent value of blank units.

## Segregated Witnesses

The Bitcoin protocol has long suffered from an issue known as transaction malleability.[5] The information used to generate the transaction ID can, in certain circumstances, be modified by a network node to generate a different transaction ID without impacting the sender, recipient, or amount transacted. Note that this just changes the hash; the output of the transaction remains the same and the bitcoins will go to their intended recipient. However this does mean that, for instance, it is not safe to accept a chain of unconfirmed transactions under any circumstance because the later transactions will depend on the hashes of the previous transactions, and those hashes can be changed until they are confirmed in a block (and potentially even after a confirmation if the blockchain is reorganized). In addition, clients must always actively scan for transactions to them; assuming a transaction output (txout) exists because the client that created it previously is unsafe. Attacks exploiting this vulnerability are speculated to be a possible cause for the MtGox failure.[6]

Recently Bitcoin was updated with a feature called Segregated Witnesses. This is a proposal to allow transaction-producing software to separate (segregate) transaction signatures (witnesses) from the rest of the data in a transaction, and to allow miners to place those witnesses outside of the traditional block structure. This provides two immediate benefits:

> **Elimination of Malleability:** Segregating the witness allows both existing software and upgraded software that receives transactions to calculate the transaction identifier (txid) of segregated transactions without referencing the witness. This solves all known cases of unwanted third-party transaction malleability, which is a problem that makes programming Bitcoin wallet software more difficult and which seriously complicates the design of smart contracts for Bitcoin.

> **Capacity Increase:** Moving witness data outside of the traditional block structure (but still inside a new-style block structure) means new-style blocks can hold more data than older-style blocks, allowing a modest increase to the amount of transaction data that can fit in a block.

Segregating witnesses also simplifies the ability to add new features to Bitcoin and improves the efficiency of full nodes, which provides long-term benefits that will be described in more detail later in this document.

Equibit, being based on the Bitcoin protocol, would also suffer from the same transaction malleability issue if a fix is not implemented. Segregated witnesses provides an informative example of a solution to this problem, however a more elegant solution is possible since we do not have to worry about a hard fork from the Equibit blockchain while it has yet to be publicly launched.

Splitting the merkle tree into two separate root hashes can help create a much smaller blockchain. One tree being the basic equibit transactions and another being the entire transaction with issuance data and signatures. This allows full audit capability for blockchains that maintain both but also allows for a much smaller blockchain where older issuance data and signatures can be ignored, but the history of the transactions can still be confirmed.

# Transaction Fees

Transaction fees in Equibit will function in the same way as the Bitcoin network, with participating nodes receiving blank equibits. As with Bitcoin, a larger transaction fee will encourage more miners and faster processing of transactions thus incentivising users to pay proportional to their need.

# Cross-Blockchain Trades

We use the Atomic Cross-Chain (ACC) trading protocol, as designed by TierNolan, to ensure the safety of a trade between buyer and seller.[7] In an atomic trade, both parties to an exchange must complete their side of the transaction and any of the parties may back out of the transaction at any time prior to completing the process. ACC requires 4 separate transactions, one on each chain (Equibit and Bitcoin) to transfer value, and 2 additional "refund" transactions that are timelocked to the future and are only posted to the blockchain in the exceptional cases where one of the parties desires to cancel the transfer.

The following algorithm describes the ACC process. Note that transactions are exchanged off-blockchain between buyer and seller until the transactions are submitted to the network.

1. Buyer picks a random number x
2. Buyer creates a bitcoin transaction TX1:
    Pay BTC to <Seller's public key> only if
        i. x for H(x) known and signed by Seller or
        ii. Transaction signed by Buyer & Seller
3. Buyer creates a refund bitcoin transaction TX2:
    Pay BTC from TX1 to <Buyer's public key>,
        i. timelocked 48 hours in the future, signed by Buyer
4. Buyer sends bitcoin TX2 to Seller
5. Seller signs bitcoin TX2 and returns to Buyer
6. Buyer submits TX1 to the bitcoin network -- *this is the payment blockchain transaction*
7. Seller creates equibit transaction TX3:
    Transfer equibits to <Buyer's public key> only if
        i. x for H(x) known and signed by Buyer or
        ii. Transaction signed by Buyer & Seller
8. Seller creates a refund equibit transaction TX4:
    Transfer equibits from TX3 to <Seller's public key>,
        i. timelocked 24 hours in the future, signed by Seller
9. Seller sends equibit TX4 to Buyer
10. Buyer signs equibit TX4 and sends back to Seller
11. Seller submits TX3 to the equibit network -- *this is the equity blockchain transaction*
12. Buyer transfers equibits from TX3 to new Buyer address, revealing value of secret x
13. Seller transfers bitcoins from TX1, to new Seller address, using revealed value of secret x

This is atomic (with timeout). If the process is halted, it can be reversed no matter when it is stopped:

**Before Step 6:** Nothing public has been broadcast, so nothing happens

**Between 6 & 11:** Buyer can use refund transaction TX2 after 48 hours to get their money back

**Between 11 & 12:** Seller can use refund equibit transaction TX4 after 24 hours

**After 12:** Transaction is completed in 2 steps:
 - Buyer must transfer their equibits within 24 hours or Seller can claim the refund
 - Seller must transfer their bitcoins within 48 hours or Buyer can claim the refund

# Network Communications

There are a number of types of communications that have to happen in an equity market. In order to facilitate these a secure, peer-to-peer communications system is needed in order to maintain the decentralized nature of Equibit. Equibit thus contains its own communications system using Secure Sockets Layer (SSL) and Elliptic Curve Cryptography (ECC) digital signatures.

These communications all happen off-blockchain and are stored locally on the network nodes involved in the discussion. Putting them on the blockchain would mean that every message ever sent over the system would be kept in a perpetual record that would cause the blockchain file size to grow rapidly and unpredictably. However the system relies on the same public and private keys that the Equibit blockchain does, making its use much easier than an entirely separate system with its own addresses.

The messaging protocol uses a Proof-of-Work (PoW) algorithm in order to prevent spamming of the network. Messages are encrypted, signed, and broadcasted to nodes, but using the correspondents' keys to ensure privacy. When a node receives a message, it first validates the PoW, and then looks at its local wallet(s) to see if it has a corresponding key to decrypt the message; once a message is decrypted, the receiving node sends a confirmation message to the sender's address. Equibit allows for three basic message types:

> **Direct Message:** These are private messages sent from one Equibit node to another. This allows network actors to bargain and exchange other information in a secure manner.
>
> **Group Message:** These are similar to direct messages, but are sent to an entire group of recipients who can all join in on the conversation and see what everyone is saying. These message types are most useful for issuers, who can create message groups with their investors and send notices, polls, and other important information to them as well as collect their responses.
>
> **Public Message:** Unlike the previous two types of messages, which are encrypted and only readable by the intended recipient's, public messages can be read by everyone. As such they are limited to certain message types such as bid/ask messages for the order book, payment addresses, proxy designations, and trading passport revocations.

When a new node joins the network, or rejoins after an extended period offline it will need to be brought up to date by its peers on the relevant activity that has occurred in its absence. For private and group messages we propose that nodes store these communications for two days and then delete them, except in the case of polls where they would be kept until the closing date. A node could choose, for its own reasons, to keep copies of these messages for longer periods as well. If a node is offline for more than two days, the sending node will notice that it never received a delivery confirmation and rebroadcasts the message after an additional two days. It will continue to rebroadcast the message, with exponential backoff, forever. In this respect private and group messages for offline recipients follows the same behavior as the Bitmessage protocol.[8]

Messages pertaining to the order book will be kept until the order reaches its expiry date since expired offers are no longer valid and able to be accepted by anyone on the network. In the case of payment addresses and proxy designations, these do not expire unless revoked or superseded. All nodes then, must have a complete set of payment addresses in order to distribute earnings and current proxies in order to correctly route any polls. Trading passport revocations are an extremely important message type, thus all revocations since the last one the node received must be collected from its peers.

## Order Book

The Equibit system facilitates the bid/ask process between potential sellers and buyers. These offers are done off-blockchain using the messaging system. The process can be initiated by either the seller (ask) or the buyer (bid) of an equity.

The offer terms are packaged and broadcasted to any users of the system. As with other message types within the system, the sender has to complete the required PoW to send the message.

Offer messages can be safely ignored, responded to (with counter-offer terms), or can be used as the basis to initiate a trade. Prior to initiating a trade, the seller may be required to validate the buyer's credentials. This process is described in the Trading Passports section below. The transfer of equibit and bitcoin between buyer and seller follows the protocol described in the Cross-Blockchain Trades section above.

Historical order book data can be compiled by any user of the system, but is not recorded in the blockchain itself. The blockchain is reserved to capture actual sale transactions and ownership of the equibits.

## Earnings Distributions

In order to distribute earnings issuers will need to select the date and time of record which the system will use to determine who all the investors are at that given date. An amount of bitcoins will also be specified as the gross amount of the dividend. After confirming the date and amounts the system will allocate the gross dividend across the relative share of authorized equibits held at each address and initiate bitcoin transfers to the payment address specified by each one.

Knowing which Bitcoin address corresponds to the Equibit address where the shares reside then becomes of paramount importance for the entire system to function and maintain a positive user experience. To create these associations, investors can create a public message that is broadcasted to the network designating the various addresses and cryptocurrencies they can be paid in.

## Polls and Proxies

Investor polling allows an issuer to communicate with anyone that owns any quantity of the issuer's authorized equibits. As with other types of messages, votes and polls are done off-blockchain. The polling messages contain a question and list of possible answers. The messages follow a standard JavaScript Object Notation (JSON) format that the issuer encrypts using the public keys of all current owners of the issuer's authorized equibits.

The Equibit nodes connected to the intended recipients' wallets are then able to validate the PoW, decrypt the poll message, query the user for answers to the poll's questions, and encrypt a response back to the issuer.

The issuer's equibit node collects the multiplicity of messages received from the network, validates the sender's signatures, and aggregates the responses to the poll. A sample poll message is illustrated below:

```json
{
    "eqbPoll": {
        "pollGUID": <poll public address>,
        "issuerID": <issuer public address>,
        "description": "Question for our shareholders",
        "closePoll": "yes",
        "closeDate": "2017-03-30T00:42:00",
        "questions": {
            "question": {
                "text": "Do you like polls?",
                "multipleChoice": "no",
                "answers": {
                    "answer": [{
                        "text": "Yes",
                        "value": "0",
                    }, {
                        "text": "No",
                        "value": "0",
                    }]
                }
            }
        }
    }
}
```

Oftentimes, investors will designate an individual or organization to speak on their behalf when it comes to polls. This typically happens with publicly traded companies, investment funds, or other kinds of managed trading accounts where the investor is simply concerned with returns and is not actively engaged in the operations of the companies they invest in.

In order to facilitate the most flexibility when creating proxies and handling their communications, three possible proxy types have been included in the Equibit platform:

**General:** Proxy can vote on behalf of the investor on any poll.

**Specific Issuer:** Proxy can vote on behalf of the investor on any poll from a specific issuing address.

**Specific Poll:** Proxy can vote on behalf of the investor on a single poll.

Designating a proxy is much like setting up a forwarding address for emails. In Equibit, designating a proxy is simply another type of public message that informs all the users who should receive a copy of certain messages sent to a particular investor, and under what conditions a response from the proxy would be accepted.

Because it's possible for an investor to have more than one proxy, it's possible that those proxies may cast a vote on the same poll. In order to avoid confusion we can assign authorities to the various types of proxies and only the vote coming from the highest authority will be tabulated. Though the other votes cast by the other proxies will still become part of the poll's records.

A proxy for a specific poll carries the highest authority of any proxy due to the fact that the investor would not bother designating one if they were not certain they wanted that proxy's vote to be the one that was tabulated. This is followed by a proxy for a specific issuer for similar reasons when considered relative to a general proxy, which has the lowest authority. A vote cast by the investor themselves will always be tabulated ahead of any proxy vote.

## Trading Passports

Many jurisdictions differentiate between publicly available, and private shares. In order to make their securities available to the public, issuers are required to produce a document known as a prospectus. The contents of these documents are defined in securities legislation.[9]

Due to the high cost of producing these disclosures, and the ongoing reporting requirements imposed on public issuers, many firms opt to make their offerings under one or more possible exemptions, a popular one being the Accredited Investor (AI) exemption. Accredited Investors are individuals or organizations with certain defining characteristics but mainly meeting minimum net worth levels.

Issuers have a positive obligation to ensure that only AI's come to own their securities when relying on this prospectus exemption. In the current, centralized control model of the equity market this is handled by the fact that updating a shareholder register requires manual intervention from the issuer or their transfer agent. Thus providing the opportunity for them to perform Know-Your-Client (KYC) and other due diligence on the buyer ensuring they're also an AI.

A blockchain-based share register presents a problem in this matter of regulatory compliance as the case is with these systems that units are transferable to any address. Thus, a solution is needed to ensure that shares issued under the AI exemption will forever remain in the ownership of other AI's as they circulate on the network without intervention or reliance on the issuers.

Trading passports are a way for issuers relying on an exemption to quickly and easily create or access communities of AI's. This eliminates the need to do due diligence on buyers on a transactional basis. The exempt distributions can circulate within these groups without the risk of them leaving and falling into the hands of unqualified investors.

This system relies on a Web of Trust (WoT), a protocol used in PGP, GnuPG and other OpenPGP-compatible systems to establish the binding between public keys. A public key becomes trusted when it is signed by the private key of another entity on the network. This results in a trading passport that can be stored by the recipient of trust and presented in a transaction where the trusted address would obtain shares from an exempt distribution.

Recalling the section on Equity Creation, Issuance and Cancellation, when an issuer authorizes new shares they have the opportunity to decide what level those restrictions will be at. Issuers with a prospectus would not normally restrict the trading of their shares as any member of the public may own them under present-day securities regulations. Any company relying on an exemption would choose from three possible levels of restriction.

Generally speaking, a restricted share is one that an investor cannot sell or transfer to other investors on their own. The only transaction permitted on a restricted share is one that sends it back to the issuer. This is how the present equity market handles transferring these shares and it is an onerous process involving lawyers, certificate verification, cancellation and re-issuance to the new investor.

By using trading passports we can create certain exceptions to this rule depending on how comfortable the issuer is and how big of a market they want to create for their shares. This is governed by the "Restriction Level" field in the authorized equibit. There are three possible settings, each making it progressively more difficult for investors to transfer the authorized equibit.

**Level 0:** No restriction. Equibit can trade freely.

**Level 1:** Equibit can be transferred to an address that is two or less degrees of trust separation from the issuer.

**Level 2:** Equibit can be transferred to an address that is one degree of trust separation from the issuer.

**Level 3:** No exceptions. Equibit can only be returned to the issuer.

When dealing with restricted shares at Level 1, trust intermediaries called "Accreditors" can be relied upon. These network actors have received the trust of many issuers and, in turn, trust many investors by doing the due diligence and KYC work that normally would rest on the shoulders of issuers. Through the WoT, Accreditors create communities of trustworthy investors where the restricted shares can flow freely, but not leave.

We can represent these degrees of trust separation simply with directional lines, indicating who trusts whom:

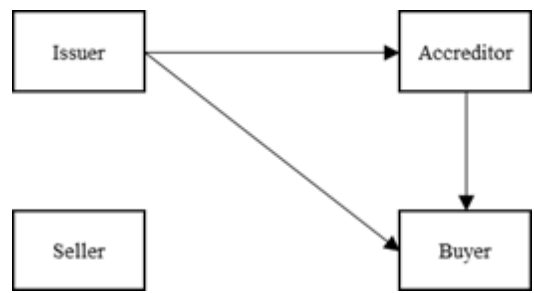

In the illustration you can clearly see the one degree of trust where the issuer has directly trusted the buyer, and the two degrees of trust where the issuer has trusted the buyer through an accreditor. For simplicity's sake we have ignored any trust lines between the issuer and the seller. In order to prevent manipulation of this system the seller is prohibited from also being an accreditor in a transaction involving two degrees of trust separation.

Proving the required number of degrees of trust exist is a simple matter of the selling node querying the purchasing node for its trading passports. If the selling node is provided with a trading passport from the issuer then the seller has what it needs to complete the transaction under both Level 1 and Level 2 restrictions. If the selling node does not receive a passport from the issuer, and the shares carry a Level 1 restriction it only then needs to query the nodes from which the buyer has received their passports to see if any of them have been trusted by the issuer. If provided with such a passport then the two degrees of trust separation have been proven and the transaction will be validated by the network. It should be noted that the issuer's node need not be involved in this process at all, thus making the network work very efficiently and removing the need for the issuer to constantly be online.

The task of being an accreditor is an important one. They perform all the KYC and additional due diligence functions that issuers can then rely on. Through accreditors, issuers can rapidly connect to pools of capital and be comfortable knowing that all these investors qualify to own their shares. Conversely, investors can gain access to companies they otherwise would not be able to own by purchasing trading passports from accreditors with many issuers behind them. Thus, a successful accreditor is in a lucrative position and could charge issuers for carrying their passports and proving the required degrees of trust separation in a trade, and charge investors for their own trading passports. These passports would necessarily have to expire after some point and since they are stored locally by the participating nodes, a public message would be broadcasted notifying the network that a particular trading passport has been invalidated.

## Conclusion

Equibit solves a current equity market need by replacing an antiquated and expensive single-trust system with a peer-to-peer network. It exponentially increases openness and forces transparency on a system that is growing in popularity among investors and issuing companies. Integrating seamlessly with the original Bitcoin protocol also allows Equibit to more easily gain traction within an already robust community and to take advantage of the growing pool of capital. Overall, we believe that Equibit will greatly advance how companies distribute equity and handle their investor relations.